\documentclass{ptephy}

\begin{document}

\title{Eigenvalues of the Breit Equation}

\author{ \name{ \fname{Yoshio} \surname{Yamaguchi}}{1} and 
 \name{\fname{Hikoya} \surname{Kasari}}{2,\ast} 
\thanks{The author Y. Y. contributed mainly to this work.}
} 

\address{
\affil{1}{Nishina  Center,  RIKEN}
\affil{2}{
Department of Physics, School of Science, 
Tokai University 259-1292 4-1-1 Kitakaname, 
Hiratsuka, Kanagawa, Japan}
\email{kasari@keyaki.cc.u-tokai.ac.jp} }

\begin{abstract}
Eigenvalues of the Breit Equation
\begin{eqnarray*}
\left [ (\vec{\alpha}_{1} \vec{p} + \beta_{1}m)_{\alpha \alpha^{\prime}} 
\delta_{\beta \beta^{\prime}} 
+ \delta_{\alpha \alpha^{\prime}} 
(-\vec{\alpha}_{2} \vec{p} + \beta_{2}M)_{\beta \beta^{\prime}} 
- \frac{e^{2}}{r} \delta_{\alpha \alpha^{\prime}} 
\delta_{\beta \beta^{\prime}}
 \right ] \Psi_{\alpha^{\prime} \beta^{\prime}} = E \Psi_{\alpha \beta},
\end{eqnarray*}
in which only the static Coulomb potential is considered, have been found.
Here the detailed discussion on the simple caces, 
$^{1}S_{0},\  m=M$ and $m \neq M$ 
is given 
deriving the exact energy eigenvalues. 
The $\alpha^2$ expansion is used to find radial wave functions.
The leading term is given by classical Coulomb wave function.
The technique used here can be applied to other cases.
\end{abstract}

\subjectindex{The eigenvalues of the Breit equation}

\maketitle

\section{Introduction and the Breit equation}
\label{sec:1}

\subsection{Introduction}
\label{sub:1-1}
The Breit equation \cite{rf:1} has traditionally \cite{rf:2} 
considered to be singular and has never been tried to solve itself.  
Instead, the Pauli approximation or a generalized Foldy-Wouthysen 
transformation has been applied to derive the effective 
Hamiltonian $H_{\hbox{eff}}$, consisting of the Schr\"{o}dinger Hamiltonian
\begin{eqnarray*}
H_{0} = -\frac{1}{2} \left ( \frac{1}{m} + \frac{1}{M} \right ) p^{2} 
- \frac{e^{2}}{r}
\end{eqnarray*}
and many other relativistic correction terms \cite{rf:2} \cite{rf:3}.  
Then perturbation method was used to evaluate the eigen-values 
(binding energies) in power series in $\alpha$ as high as possible 
(or as much as needed for experimental verification of QED).

Here we shall try to directly solve the Breit equation.  
The radial wave fundtion for the simplest case, $^{1}S$ leptonium states 
with equal masses, is given by the confluent Heun function. 
The auther is not familiar to handle this function. 
Instead, the approximate method using $\alpha^{2} \sim 10^{-4}$ 
expansion shall be used to find 
wave funcitons. 
Its leading term is given by Coulomb wave function.

However, the exact eigenvalues are found in singlet-states with equal 
$m=M$ and unequal masses $m \neq M$ cases.
We shall study the simplest case, $^{1}S$, equal lepton masses, 
$m=M$, in greater details here. 
The technique used here shall be applied to other cases, 
$^{1}(l)_l, \ ^{3}(l)_l$ and $^{3}P_0$ 
including $m \neq M$, to find exact eigenvalues.

More complicated cases, $m \neq M$ described in \S \ref{sub:2-3} below 
and triplet cases, shall be described in due course.

\subsection{Basic equation}
\label{sub:1-2}
The Breit equation for two leptons (Dirac particles), 
with charge and mass, $(-e, m)$ and $(+e,M)$, 
interacting through the (static) Coulomb potential 
$-e^{2}/r = -\alpha/r \, 
(\alpha \lower.8ex\hbox{$\stackrel{\doteq}{ .}$} 1/137)$ 
is given by
\begin{eqnarray}
\left [ (\vec{\alpha}_{1} \vec{p} + \beta_{1}m)_{\mu \nu} 
\delta_{\rho \sigma} 
+ \delta_{\mu \nu} 
(-\vec{\alpha}_{2} \vec{p} + \beta_{2}M)_{\rho \sigma} 
- \frac{e^{2}}{r} \delta_{\mu \nu} \delta_{\rho \sigma}
 \right ] \Psi_{\nu \sigma} = E \Psi_{\mu \rho}.
\label{eq:1}
\end{eqnarray}
This equation holds in the CM system.  
The total energy of the system $E$ can be expressed as
\begin{eqnarray}
E = \sqrt{m^{2} - q^{2}} + \sqrt{M^{2} - q^{2}}.
\label{eq:2}
\end{eqnarray}
The momentum operator in (\ref{eq:1}) can be written as
\begin{eqnarray}
\vec{p} = \frac{1}{i} \vec{\nabla} = 
\frac{1}{i} \frac{\partial}{\partial \vec{r}}\, ,
\label{eq:3}
\end{eqnarray}
introducing the operator $\vec{r}$ canonically conjugate to $\vec{p}$.  
Then the (static) Coulomb potential 
(contribution form one longitudinal photon exchange between two oppositely 
charged point Dirac particles) can be written as 
$-e^{2}/r = -\alpha/r$.  $\vec{\alpha}_{j}, \beta_{j} (j=1,2)$ 
are the usual $4 \times 4$ Dirac matrices for particle 1 $(-e,m)$ 
and 2 $(+e,M)$, respectively.  
$\Psi_{\alpha \beta}$ is $4 \times 4$ Dirac spinor wave function.

We shall try to solve
\begin{eqnarray}
\left [(\vec{\alpha}_{1} \frac{1}{i} \vec{\nabla} + \beta_{1}m) + 
(- \vec{\alpha}_{2} \frac{1}{i} \vec{\nabla} + \beta_{2}M)
- \frac{\alpha}{r} \right ]
\Psi = E \Psi,
\label{eq:4}
\end{eqnarray}
and find the eigen-value(s)
\addtocounter{equation}{-3}
\begin{eqnarray}
E = \sqrt{m^{2} - q^{2}} + \sqrt{M^{2} - q^{2}},
\label{eq:2b}
\end{eqnarray}
which is specified by discrete real value(s) $q_{n}$.

The two particle system can be classified by the total angular momentum 
and its z-component and parity.  
In so doing it is sufficient to specify spin angular parts of 
the large-large components $\Psi_{\alpha \beta} \ (\alpha, \beta = 1,2)$ to be
\addtocounter{equation}{2}
\begin{eqnarray}
\begin{array}{lll}
& ^{1} (l)_{l}, & (l=0,1,2, \cdots)\\
& ^{3}(l-1)_{l} + ^{3}(l+1)_{l}, & (l=1,2, \cdots)\\
\hbox{and} & ^{3}(l)_{l}, & (l=1,2, \cdots)
\end{array}
\nonumber
\end{eqnarray}
where $l$ is the quantum number for the orbital angular momentum.  
Spin-angular parts of other (large-small, small-large, small-small) 
components of $\Psi_{\alpha \beta}$ are completely fixed by those of 
the large-large components.

We shall describe the singlet case in \S\ref{sec:2} in detail.  
Other cases can be handled similarly.

\section{The Singlet Cases}
\label{sec:2}
\subsection{$^{1}(l)_{l}$, general}
\label{sub:2-1}
We shall first discuss the simplest case of $^{1}(l)_{l}$.  
$\Psi_{\alpha \beta}$ can be taken as follows:
\begin{eqnarray}
\begin{array}{l|c}
\hline
\begin{array}{ll}
\alpha & \beta
\end{array}
& \Psi_{\alpha \beta} \\
\hline
\begin{array}{ll}
1 & 1 \\
2 & 2
\end{array}
&
F(r) | ^{1}(l)_{l}> \\
\hline
\begin{array}{ll}
1 & 3 \\
2 & 4
\end{array}
&
i\{G(r) | ^{3}(l+1)_{l}> + \tilde{G}(r) | ^{3}(l-1)_{l}> \}\\
\hline
\begin{array}{ll}
3 & 1 \\
4 & 2
\end{array}
&
i\{H(r) | ^{3}(l+1)_{l}> + \tilde{H}(r) | ^{3}(l-1)_{l}> \}\\
\hline
\begin{array}{ll}
3 & 3 \\
4 & 4
\end{array}
&
K(r) | ^{1}(l)_{l}>
\end{array}
\label{eq:5}
\end{eqnarray}
where $|^{1}(l)_{l}>$ and $|^{3}(l \pm 1)_{l}>$ are 
normalized spin-angular wave functions 
($z$-component of the total angular momentum has been omitted).  
The Breit equation (\ref{eq:4}) in \S\ref{sec:1} gives the following 
set of differential equations for the radial wave functions 
$F(r), K(r), G(r), \tilde{G}(r), H(r)$ and $\tilde{H}(r)$:
\begin{eqnarray}
\lefteqn{ \left( E + \frac{\alpha}{r} - m - M \right ) F} \nonumber \\
& & \qquad = \sqrt{\frac{l+1}{2l+1}} 
\left (\frac{d}{dr} + \frac{l+2}{r} \right ) (G+H) 
- \sqrt{\frac{l}{2l+1}} \left (\frac{d}{dr} 
- \frac{l-1}{r} \right ) (\tilde{G} - \tilde{H}), \nonumber \\
\lefteqn{\sqrt{l} \left (\frac{d}{dr} + \frac{l+2}{r} \right ) (G-H) 
+ \sqrt{l+1} \left (\frac{d}{dr} - \frac{l-1}{r} \right ) 
(\tilde{G} - \tilde{H}) = 0,} \nonumber \\
\lefteqn{ \left [ E + \frac{\alpha}{r} - (m+M) \right ] F 
= \left [ E + \frac{\alpha}{r} + (m+M) \right ] K,} \nonumber \\
\lefteqn{ \left ( E+\frac{\alpha}{r} + m-M \right ) H 
= - \sqrt{\frac{l+1}{2l+1}}
 \left (\frac{d}{dr} - \frac{l}{r} \right ) (F+K)} \nonumber \\
& & \qquad = \left ( E+\frac{\alpha}{r} - m+M \right ) G, \nonumber \\
\lefteqn{ \left ( E+\frac{\alpha}{r}+m-M \right ) \tilde{H} = 
\sqrt{\frac{l}{2l+1}} \left (\frac{d}{dr}+\frac{l+1}{r} \right )(F+K)} 
\nonumber \\
& & \qquad = \left ( E + \frac{\alpha}{r} - m+M \right ) \tilde{G}.
\label{eq:6}
\end{eqnarray}
Therefore, we have the relations:
\begin{eqnarray}
\left ( E + \frac{\alpha}{r} +m- M \right )  H 
= \left ( E + \frac{\alpha}{r} -m+ M \right ) G, 
\nonumber \\
\left ( E + \frac{\alpha}{r} +m- M \right ) \tilde{H} 
= \left ( E + \frac{\alpha}{r} -m+ M \right) \tilde{G}.
\nonumber
\end{eqnarray}

For later convenience, we introduce the dimensionless quantities:
\begin{eqnarray}
\rho = 2qr,\ \ \ \ \ \ \ \ \ \ \ \ \ \ \ \ \ \ \ \ \   \nonumber \\
\nonumber \\ 
y = \frac{E}{2 \alpha q} 
= \frac{\sqrt{M^{2} - q^{2}}+ \sqrt{m^{2} - q^{2}}}{2 \alpha q} \;\;\; (>0), 
\nonumber \\
\lambda = \frac{M+m+E}{2 \alpha q} \;\;\; (>0), \nonumber \\
\nu = \frac{M+m-E}{2 \alpha q} \;\;\; (>0).
\label{eq:7}
\end{eqnarray}

If the eigen-value $E$ is a proper relativistic generalization of 
the Schr\"{o}dinger case (as we shall see it is the case), 
$y$ and $\lambda$ are large quantities of the order of $1/\alpha^{2}$,
 while $\nu$ is the order of unity.

For $\rho \rightarrow \infty (r \rightarrow \infty)$ all radial wave function 
should behave like
\begin{eqnarray}
F(\rho) \rightarrow \rho^{n} \{1+O(1/\rho) \} e^{-\rho/2}.
\label{eq:8}
\end{eqnarray}
This is because $e^{-\rho/2}=e^{-qr}$ and 
$E=\sqrt{m^{2} - q^{2}} + \sqrt{M^{2}-q^{2}}$.  
Notice that if we consider $E$, the total CM energy 
$\sqrt{m^{2}+p^{2}} + \sqrt{M^{2}+p^{2}}$, to be larger than 
$M+m$, we would expect that the radial wave functions should contain 
a factor $e^{ipr}$.
If we analitically continue $p$ to imaginary values 
we get the factor $e^{-qr}$ and the energy expression (\ref{eq:2})
for bound states.

The singlet case becomes very simple when two masses are equal $M=m$.  
First
\begin{eqnarray}
\left.
\begin{array}{l}
H=G, \;\;\; \tilde{H}=\tilde{G}, \\
 \\ 
\displaystyle{ K = \frac{1-\nu \rho}{1+ \lambda \rho}F }
\end{array}
\right \}
\label{eq:9}
\end{eqnarray}
Then, the combination 
\begin{eqnarray}
F+K = \tilde{h}(\rho)e^{-\rho/2}
\label{eq:10}
\end{eqnarray}
obeys the differential equation
\begin{eqnarray*}
\lefteqn{
\tilde{h}^{\prime \prime} + \tilde{h}^{\prime}
\left \{-1 + \frac{2}{\rho} + \frac{1}{\rho (1+y\rho)} \right \}
} \nonumber \\
& & \qquad 
+ \tilde{h} \left [ \left (\frac{\alpha^{2}y}{2} -1 \right ) \frac{1}{\rho} 
+ \frac{\alpha^{2}}{4} \frac{1}{\rho^{2}}
-\frac{l (l+1)}{\rho^{2}}
- \frac{1}{2 \rho (1+y\rho)} \right ] = 0.
\end{eqnarray*}

\subsection{$^{1}S_{0}$, $M=m$}
\label{sub:2-2}
The $^{1}S_{0}$, i.e., $l=0$, case we have the equation for $\tilde{h}$
\begin{eqnarray}
\lefteqn{
\tilde{h}^{\prime \prime} + \tilde{h}^{\prime}
\left \{-1 + \frac{2}{\rho} + \frac{1}{\rho (1+y\rho)} \right \}
} \nonumber \\
& & \qquad
+ \tilde{h} \left [ \left (\frac{\alpha^{2}y}{2} -1 \right ) \frac{1}{\rho} 
+ \frac{\alpha^{2}}{4} \frac{1}{\rho^{2}}
- \frac{1}{2 \rho (1+y\rho)} \right ] = 0.
\label{eq:11}
\end{eqnarray}

If we ignore $\alpha^{2}$ and higher order terms in 
(\ref{eq:11}) and notice that $\alpha^{2}y$ is of 
the order of unity,
we find
\begin{eqnarray}
\tilde{h}^{\prime \prime} + \tilde{h}^{\prime} 
\left \{-1 + \frac{2}{\rho} \right \}
+ \tilde{h} \left (\frac{\alpha^{2}y}{2} -1 \right ) \frac{1}{\rho} = 0,
\label{eq:12}
\end{eqnarray}
which is precisely the Schr\"{o}dinger equation for our issue and 
$\tilde{h}$ in this approximation is given by 
$F (-\frac{\alpha^{2}y}{2} + 1, 2, \rho) = F(1-n, 2, \rho)$ 
apart from the normalization constant.  
$\frac{\alpha^{2}y}{2} -1$ must be equal to an integer $n-1=0,1,2, \cdots$.

(\ref{eq:11}) shows that $\rho=0$ is the regular singular point while 
$\rho = \infty$ is an irregular singular point.

Near $\rho = 0 \;\; \tilde{h} (\rho)$ can be expressed as 
\begin{eqnarray}
\left.
\begin{array}{lr}
& 
\displaystyle{
\tilde{h}(\rho) = \rho^{s}h(\rho) 
}
, \\
  \\ 
\hbox{where } & 
\displaystyle{
s= -1 + \sqrt{1 - \frac{\alpha^{2}}{4}}.
}
\end{array}
\right \}
\label{eq:13}
\end{eqnarray}
Another solution $s = -1-\sqrt{1 - \frac{\alpha^{2}}{4}}$ 
is unacceptable from the square integrability of the wave function 
$\Psi_{\alpha \beta}$.  
$h(\rho)$ obeys the equation:
\begin{eqnarray}
\lefteqn{
h^{\prime \prime} + h^{\prime} 
\left \{-1 + \frac{2+2s}{\rho} + \frac{1}{\rho (1+y\rho)} \right \}
} \nonumber \\
& & \qquad + h 
\left \{ \left (\frac{\alpha^{2}y}{2} - 1 - s \right ) 
\frac{1}{\rho} - \left (\frac{1}{2} + sy \right ) 
\frac{1}{\rho (1+y\rho)} \right \} = 0.
\label{eq:14}
\end{eqnarray}
$h(\rho)$ can be expanded in a Taylor series near $\rho=0$:
\begin{eqnarray}
\lefteqn{h(\rho) = \sum_{n=0}^{\infty} h_{n} \rho^{n},}
\label{eq:15}
\\
\lefteqn{h_{1} = \frac{1}{3+2s} \left \{ - \frac{\alpha^{2}y}{2} 
+ 1 + s + \frac{1}{2} + sy \right \} h_{0},}
\nonumber \\
& & \qquad (n+1) (n+3+2s) \frac{1}{y} h_{n+1}
\nonumber \\
& & \qquad + \left \{ n (n+1+2s) - \frac{1}{y} 
(n - \frac{\alpha^{2}y}{2} +1+s+\frac{1}{2} + sy) \right \} h_{n} 
\nonumber \\
& & \qquad - \left \{n-1- \frac{\alpha^{2}y}{2} +1+s \right \} h_{n-1} = 0.
\label{eq:16}
\end{eqnarray}
It can be shown that $h(\rho)$ can not be a polynomial:
\begin{eqnarray*}
h(\rho) = \sum _{n=0}^{N} h_{n} \rho^{n} \;\;\; (N < \infty)
\end{eqnarray*}
from the recurrence formulae (\ref{eq:16}).  
So the sum in equation (\ref{eq:15}) must extend to $n=\infty$, 
a sharp difference from the Schr\"{o}dinger case.
Note that the expansion (\ref{eq:15}) holds only the tiny region 
$\rho < 1/y \sim O(\alpha^2)$.

Next we discuss on $\rho = \infty$, which is an irregular singular point.  
Assuming the form
\begin{eqnarray}
h(\rho) = e^{\lambda \rho} \rho^{k} \left [c_{0} 
+ \frac{c_{1}}{\rho} + \frac{c_{2}}{\rho_{2}} + \cdots \right ] \;\;\; 
\hbox{at } \rho \rightarrow \infty
\label{eq:17}
\end{eqnarray}
where $\lambda$, $k$, $c_{n}$ are constant.  
Introducing (\ref{eq:17}) into the differential equation (\ref{eq:14}), 
we find $\lambda$ is either 0 or 1 while $k$ is undermined.  
From the square-integrability of the wave function we must choose 
$\lambda$ to be 0 and $k < \infty$.

$k$ turns out to be
\begin{eqnarray}
k = \frac{\alpha^2 y}{2} - 1 - s
\label{eq:18}
\end{eqnarray}
appeared in (\ref{eq:14}), whose value is not fixed from the situation 
at $\rho \rightarrow \infty$.
Further discussion will be given in \S 2.2.4.

\subsubsection{Detailed discussion on $^{1}S_0, \ m=M$}

Note again 
\begin{eqnarray*}
F + K = e^{-\rho/2} \rho^s h(\rho).
\end{eqnarray*}
$h(\rho)$ obeys the diffrential equation, (\ref{eq:14}),
\begin{eqnarray}
\left.
\begin{array}{ll}
& 
\displaystyle{
\frac{d^2 h}{d \rho^2} +  \frac{dh}{d\rho} 
\left \{-1+\frac{\gamma}{\rho} \right \} + 
h \left \{ \frac{N}{\rho} \right \}
= 
\left \{ - \frac{1}{\rho (1 + y\rho)} 
\left ( \frac{d}{d\rho} - \delta \right ) \right \} h,
}
 \\ 
 & \\
\hbox{where } & \gamma = 2 + 2s, \ 
s= -1 + \sqrt{1 - \frac{\alpha^{2}}{4}}, \\ 
 & \\ 
& \delta = \frac{1}{2} + sy, \\ 
 & \\ 
& N = \frac{\alpha^2 y}{2} -1 -s, \\ 
 & \\ 
& \rho = 2qr, \\ 
 & \\ 
& y = \frac{2 \sqrt{m^2 - q^2}}{2 \alpha q} = \frac{E}{2 \alpha q},
\end{array}
\right \}
\label{eq:19}
\end{eqnarray}
$\alpha^2 y$ and $sy$ are of the order of unity $O(1)$.

The RHS of equation (\ref{eq:19}) is of the order of $1/y$ or 
$O(\alpha^2) \simeq 10^{-4}$.
As the first step we shall ignore the RHS of (\ref{eq:19}). 
Then $h$ can immediately solved: 
\begin{eqnarray}
h \ 
\lower.8ex\hbox{$\stackrel{\doteq}{ .}$} \ 
F(-N, \gamma, \rho).
\label{eq:20}
\end{eqnarray}
Normalizability of the radial wave function requires $N$ equals to 
positive integer:
\begin{eqnarray}
N = \frac{\alpha^2 y}{2} - 1 - s = n -1, 
\ \ \ \  n = 1, 2, 3, \cdots.
\label{eq:21}
\end{eqnarray}
$n$ is the pricipal quantum number. 
(\ref{eq:20}) is the relativistic extension form of the non-relativisitic 
Coulomb wave function
\begin{eqnarray}
\begin{array}{l}
F(-N, \gamma, \rho) \ \ \ 
\left \{
\begin{array}{rl}
N = \frac{\alpha^2 y}{2} -1 -s & \rightarrow n - 1, \\ 
 & \\ 
\gamma = 2 + 2 s & \rightarrow 2.
\end{array}
\right.
\end{array}
\label{eq:22}
\end{eqnarray}
The energy eigenvalues derived from the condition 
(\ref{eq:21}) are
\begin{eqnarray}
E_n &=& 2 \sqrt{m^2 - q^2} 
= 2 m \sqrt{1- \frac{1}{1 + (\alpha y)^2}}  \nonumber \\
 &=& 2 m \sqrt{1- \frac{\alpha^2}{ (\alpha^2 y)^2 + \alpha^2}} \nonumber \\ 
 &=& 2 m \sqrt{1- \frac{\alpha^2}{ \{2(n+s)\}^2 + \alpha^2}}.
\label{eq:23}
\end{eqnarray}
When $n=1$, 
\begin{eqnarray}
E_1= 2m \sqrt{1 - \frac{\alpha^2}{4}}.
\label{eq:24}
\end{eqnarray}

The approximate solution $F(-N, \gamma, \rho)$ is correct 
ignoring terms of the order 
$O(1/y) \simeq \alpha^2 \simeq 10^{-4}$. 
We write 
\begin{eqnarray}
h(\rho) = F(-N, \gamma, \rho) + f(\rho). 
\label{eq:25}
\end{eqnarray}

When $n=1$, $F(0, \gamma, \rho) = 1$,
\begin{eqnarray}
f(0) = 0.
\label{eq:26}
\end{eqnarray}

We shall introduce
\begin{eqnarray*}
h = F + f,
\end{eqnarray*}
into the eq. (\ref{eq:19}), then we find 
\begin{eqnarray}
\frac{d^2 f}{d \rho^2} + 
\frac{d f}{d \rho} \left \{ -1 + \frac{\gamma}{\rho} \right \} + 
f \frac{N}{\rho} = 
\left \{ 
  - \frac{1}{\rho (1 + y \rho) }
  \left ( \frac{d}{d \rho} - \delta \right )
\right \} (F + f),  
\label{eq:27}
\end{eqnarray}
whose solution gives $h=F+f$.

Or, alternatively, we may use an approximate method: 
$F$ is the order of unity, while $f(\rho)$ is the order of 
$O(1/y) \simeq O(\alpha^2)$, so that $f$ in the RHS of (\ref{eq:27}) 
can be ignored. 
This equation gives immediately the solution 
\begin{eqnarray*}
f(\rho) = \int^{\rho}_{0} \frac{e^{\rho}d\rho}{(F)^2 \rho^{\gamma}} 
\int^{\rho} e^{-\sigma} F(-N, \gamma, \sigma) \sigma^{\gamma} 
d \sigma 
\left [ -\frac{1}{\sigma(1+y \sigma)} 
        \left \{ \frac{d}{d \sigma} - \delta \right \} F
\right ].
\end{eqnarray*}
It is difficult to perform integretion here, 
but computer shall easily do the jobs.

The equation (\ref{eq:21}) give the exact eigenvalues for our problem. 
The reason is as follows. 
When $\rho \rightarrow \infty$, the function $h(\rho) = F(\rho) + f(\rho)$ 
behaves like 
(see \S\ref{subsub:2-2-4} )
\begin{eqnarray*}
h(\rho) = \rho^{\beta} \left( c_0 + \frac{c_1}{\rho} + \cdots \right), 
\ \ \ \ \ \  \beta = N
\end{eqnarray*}
as will be described in \S\ref{subsub:2-2-4}
$F(\rho)$ contributed to $\rho^{\beta}$, while $f(\rho)$ does not, 
i.e., 
\begin{eqnarray*}
\frac{f(\rho)}{\rho^{\beta}} \simeq O \left( \frac{1}{y \rho} \right), 
\ \ \ \ \ \ \ \hbox{\rm{when}} \ \ \rho \rightarrow \infty.
\end{eqnarray*}
Therefore (\ref{eq:21}) gives the exact eigenvalues. 

To find $f(\rho)$, we may use $1/y$ expansion, or $\alpha^2$ expansion, 
since $y$ is of the order of $1/\alpha^2$. 

Introducing 
\begin{eqnarray*}
h(\rho) &=& \sum_{\nu = 0}^{\infty} f^{(\nu)} (\rho), \\ 
f^{(0)} (\rho) &=& F(\rho) = F(-N, \gamma; \rho), \ \ \ \ 
N=n-1, \\ 
f(\rho) &=& \sum_{\nu=1}^{\infty} f^{(\nu)} (\rho), 
\end{eqnarray*}
where $f^{\nu} (\rho)$ is of the order of 
$(1/y)^{\nu} \sim O(\alpha^{2 \nu})$, and the differential operators
\begin{eqnarray*}
D^{(0)} &=& \frac{d^2}{d \rho^2} + 
            \left(-1 + \frac{\gamma}{\rho} \right) \frac{d}{d \rho}
            + \frac{N}{\rho}, \\ 
D^{(1)} &=& - \frac{1}{\rho(1+y \rho)} 
               \left( \frac{d}{d \rho} - \delta \right),
\end{eqnarray*}
where $D^{(\nu)}$ is of the order of $(1/y)^{\nu}$ 
(here $\nu= 0$, and $1$), the differential equation (\ref{eq:19})
will give the following serries of equations 
\begin{eqnarray*}
D^{(0)} f^{(0)} &=& 0, \\ 
D^{(0)} f^{(\nu)} (\rho) &=& D^{(1)} f^{(\nu -1)}(\rho), \ \ \ 
(\nu = 1,2,\cdots).
\end{eqnarray*}

It is easy to see that 
\begin{eqnarray*}
f^{(0)} (\rho) + f^{(1)} (\rho) = F (\rho) + f^{(1)} (\rho)
\end{eqnarray*}
shall be good if the terms of the order of $\alpha^4 \sim 10^{-8}$ 
are ignored.

These techniques can be applied to other singlet and triplet cases 
with $m \neq M$.

\subsubsection{$f(\rho)$ for the ground state 
of $^{1}S_0$, $m=M$, near $\rho=0$ ($\rho < 1/y$)}

The function $h(\rho)$ for $n=1$ is given by 
\begin{eqnarray}
\begin{array}{rl}
h(\rho) &= F(0, \gamma, \rho) + f (\rho) = 1 + f(\rho), \\
 & \\ 
f(0) &= 0.
\end{array}
\label{eq:28}
\end{eqnarray}
The Taylor expansion of $f(\rho)$ near $\rho=0$ is given by 
\begin{eqnarray*}
f(\rho) = \sum^{\infty}_{n=1} f_n \ \rho^n
\end{eqnarray*}
$f_n$'s are equal to $h_n (n \geq 1)$ and given by (\ref{eq:16}).
Noting that
\begin{eqnarray*}
\begin{array}{rl}
s \ 
\lower.8ex\hbox{$\stackrel{\doteq}{ .}$}
- \frac{\alpha^2}{8}, 
 &
\alpha^2 y \ 
\lower.8ex\hbox{$\stackrel{\doteq}{ .}$}
\ 
2,  \\
 & \\ 
sy \ 
\lower.8ex\hbox{$\stackrel{\doteq}{ .}$}
- \frac{1}{4}, 
 &
\delta \ 
\lower.8ex\hbox{$\stackrel{\doteq}{ .}$}
\ 
\frac{1}{4},  \\
 & \\ 
N \ 
\lower.8ex\hbox{$\stackrel{\doteq}{ .}$}
\  0,
 &
\gamma \ 
\lower.8ex\hbox{$\stackrel{\doteq}{ .}$}
\ 
2,
\end{array}
\end{eqnarray*}
one finds 
\begin{eqnarray*}
f_0 = 0,  \ \ 
f_1 \ 
\lower.8ex\hbox{$\stackrel{\doteq}{ .}$}
\ 
\frac{1}{12},  
\ \ 
f_2 \ 
\lower.8ex\hbox{$\stackrel{\doteq}{ .}$}
- \frac{1}{4} y f_1 = - \frac{1}{48} y, 
\end{eqnarray*}
and in general
\begin{eqnarray*}
f_{n+1} &
\lower.8ex\hbox{$\stackrel{\doteq}{ .}$}
& \frac{(-y)^n /2}{(n+1)(n+2)(n+3)}. \\ 
\end{eqnarray*}
\begin{eqnarray*}
f(\rho) &
\lower.8ex\hbox{$\stackrel{\doteq}{ .}$}
& 
\sum^{\infty}_{n=1} f_n \rho^{n} 
= \sum^{\infty}_{n=1} \frac{(-y)^{n-1}}{n (n+1) (n+2)} \frac{1}{2} \rho^n 
 \\
 & & \\ 
 & = & 
- \frac{1}{2y} 
\left [
        \frac{(1+y\rho)^2}{2 (y \rho)^2} 
        \log \left ( \frac{1}{1 + y \rho}  \right )
        + \frac{3}{4} + \frac{1}{2 y \rho}
\right ], 
\end{eqnarray*}
which holds at $-\frac{1}{y} < \rho < \frac{1}{y}$.
$f(\rho)$ is of the order of $O(1/y) = O(\alpha^2)$ and this $y f(\rho)$ 
is good within the errors of $O(1/y) = O(\alpha^2)$.
General cases $n>1$ can be treated analogously.

\subsubsection{$h(\rho)$ for the ground state of $^{1}S_0$, $m=M$, 
near $\rho = - 1/y$, mathematical curiosity}

The eq. (\ref{eq:14}) has the regular singular point at $\rho = - 1/y$. 
The regular solution near $\rho = - 1/y$ is given by 
\begin{eqnarray}
h(x) = \sum^{\infty}_{n=2} g_n x^n,
\label{eq:29}
\end{eqnarray}
where $x = \rho - \frac{1}{y}$.

Another solution near $x = 0 = \rho - \frac{1}{y}$ contains log of the 
solution (\ref{eq:29}).

The recurrence formulae for $g_n$'s are 
\begin{eqnarray}
3 g_3 = \left \{ 2 (1 + \gamma) y + 2 - \gamma \right \} g_2, 
\label{eq:30}
\end{eqnarray}
\begin{eqnarray}
-(n+1)(n-1) g_{n+1} + \left \{ n (n + \gamma y) \delta \right \} g_n 
-y (n-1-N) g_{n-1}.
\label{eq:31}
\end{eqnarray}
Ignoring $O(1/y)$ terms, one finds 
\begin{eqnarray*}
g_3 \ 
\lower.8ex\hbox{$\stackrel{\doteq}{ .}$}
\ 
2 y g_2, \ \ 
g_4 \ 
\lower.8ex\hbox{$\stackrel{\doteq}{ .}$}
\ 
\frac{3}{2} y g_3 \ 
\lower.8ex\hbox{$\stackrel{\doteq}{ .}$}
\ 
3 y^2 g_2.
\end{eqnarray*}
Taking $g_2 = 1$, 
\begin{eqnarray*}
g_{n+1} \ 
\lower.8ex\hbox{$\stackrel{\doteq}{ .}$}
\ 
\frac{ny}{(n-1)} g_n \ 
\lower.8ex\hbox{$\stackrel{\doteq}{ .}$}
\ 
n y^{n-1}, \ \ (n>1).
\end{eqnarray*}
So that $h(\rho)$ turn out to be
\begin{eqnarray*}
h(\rho) &=& \sum^{\infty}_{n=2} g_n x^n y^{n-1}
\lower.8ex\hbox{$\stackrel{\doteq}{ .}$}
\sum^{\infty}_{n=1} n x^{n+1} y^{n} \\ 
 &=& \frac{x^2 y}{1- yx} = \frac{(1+y \rho)^2}{y^2 \rho}.
\end{eqnarray*} 
This expression is valid at the unphysical range 
$- 2/y < \rho < 0$ and good within $\alpha^2$.

\subsubsection{Behavier of $h(\rho)$ at $\rho \rightarrow \infty$}
\label{subsub:2-2-4}

$\rho = \infty$ is the irreguler singular point, so that
$h(\rho)$ can be expressed in the form (\ref{eq:17}) 
as described in \S 2.2.

We shall give the recurrence formulae here for the case of 
$^{1}S, m=M$
\begin{eqnarray*}
h(\rho) &=& \rho^{\beta} 
\left ( c_{0} + \frac{c_{1}}{\rho} + \frac{c_{2}}{\rho^{2}}+ \cdots \right ), 
\\ 
c_1 &=& \left \{ -\beta (\beta + \gamma -1) 
+\frac{\delta}{y} \right \} c_0, \\ 
(n+1) y c_{n+1} &+& (n-\beta) 
\left \{ (n+1-\gamma - \beta) y -n+\delta \right \} c_{n} \\ 
&+& (n-1-\beta)(n-\gamma-1-\beta)c_{n-1} = 0,
\end{eqnarray*}
where
\begin{eqnarray*}
\gamma &=& 2 + 2s, \\ 
\delta &=& \frac{1}{2}, \\ 
\beta = N &=& \frac{\alpha^2 y}{2} -1 -s = n-1.
\end{eqnarray*}
$N$ has already been fixed in Eq. (\ref{eq:21}).

We repeat that
 $\rho^{\beta}$ is the contribution 
from $F(\rho)$ and $f(\rho) = h(\rho) - F(\rho)$ 
does not contribute to $\rho^{\beta}$ term.
Furthermore it is easy to show 
\begin{eqnarray*}
\frac{f(\rho)}{\rho^{\beta}} \sim O \left( \frac{1}{y \rho} \right)
\ \ \ \ \ \hbox{when}\ \ \rho \rightarrow \infty.
\end{eqnarray*}
This shows (\ref{eq:21}) is in fact the correct (or exact) eigenvalues.

\subsubsection{Comparison of Breit case with Dirac Coulomb solution}

For simplicity we shall discuss only ground states of $^{2}S_{1/2}$ 
(Dirac) and $^{1}S_{0}$ (Breit).

Dirac equation with Coulomb potential\cite{rf:4} reads
\begin{eqnarray}
\begin{array}{c}
\left \{ \frac{1}{i} \vec{\alpha} \vec{\nabla} + \beta m 
- \frac{\alpha}{r} \right \} \Psi = E \Psi, \\ 
 \\ 
E = \sqrt{m^2 - q^2}.
\end{array}
\label{eq:32}
\end{eqnarray}
$\Psi$ for $^{2}S_{1/2}$, spin up, can be written 
\begin{eqnarray}
\Psi = \left(
    \begin{array}{l}
     F(r) Y_{00} \frac{\sigma_{3+1}}{2} \\ 
     K(r) (\vec{\sigma} \hat{\vec{r}} ) \frac{\sigma_{3+1}}{2}
    \end{array}
\right )
    \begin{array}{l}
     s_{1/2} \ \ \hbox{spin up} \\ 
     p_{1/2} \ \ \hbox{spin up},
    \end{array}
\label{eq:33}
\end{eqnarray}
\begin{eqnarray*}
\sigma_3 = \left ( 
                   \begin{array}{rr}
                   1 & 0 \\ 
                   0 & -1
                   \end{array}
           \right ), \ \ \ 
Y_{00} (\theta, \phi) = \frac{1}{\sqrt{2 \pi}}, 
\end{eqnarray*}
\begin{eqnarray*}
\vec{\sigma} \hat{\vec{r}} = \frac{1}{r} (\vec{\sigma} \vec{r}).
\end{eqnarray*}
The differential equation for $F$ is given by 
\begin{eqnarray}
\frac{d^2 F}{dr^2} + 
   \left \{ 
     \frac{1}{E + \frac{\alpha}{r} + m} \cdot \frac{\alpha}{r^2} 
     + \frac{2}{r}
   \right \} \frac{dF}{dr} + 
  \left \{ \left ( E + \frac{\alpha}{r} \right )^2 - m^2 
  \right \} F = 0.
\label{eq:34}
\end{eqnarray}
Introducing dimensionless quantities 
\begin{eqnarray*}
\rho = 2 q r, \ \ \ y = \frac{E+m}{2 \alpha q},
\end{eqnarray*}
and write $F$ to be 
\begin{eqnarray*}
F(\rho) = e^{- \rho /2} \rho^s h(\rho), 
\end{eqnarray*}
where $s$ is introduced to remove $1/ \rho^2$ terms as usual.
Then the differential equation for $h(\rho)$ is given by 
\begin{eqnarray}
h'' &+& h' 
\left \{
 -1 + \frac{2+2s}{\rho} + \frac{1}{\rho (1 + y \rho)}
\right \} \nonumber \\
&+& h 
\left \{ \left ( \frac{\alpha E}{q} -1 -s \right ) \frac{1}{\rho} - 
\left ( \frac{1}{2} + s y \right ) \frac{1}{\rho(1 + y \rho)}
\right \} = 0,
\label{eq:35}
\end{eqnarray}
where $E= \sqrt{m^2 - q^2}$, binding energy $B=m-\sqrt{m^2 - q^2}$, 
$y = (E+m)/ 2 \alpha q$.
However $s$ is different from Breit case 
\begin{eqnarray*}
s = -1 + \sqrt{1 - \alpha^2}.
\end{eqnarray*}

The form (\ref{eq:35}) in the Dirac case looks exactly 
the same as the Breit case (\ref{eq:14}). 
Only differences are in below. 
\vspace{3mm}
\begin{center}
\begin{tabular}{|c|c|c|} \hline
 & Dirac & Breit \\ \hline
$s$ & $
\displaystyle{ 
s = -1 + \sqrt{1-\alpha^2} } $
& $
\displaystyle{
s = -1 + \sqrt{1- \frac{\alpha^2}{4}}
} 
$ \\ 
$N$ & $
\displaystyle{
\frac{\alpha E}{q} -1 -s
}
$ & $
\displaystyle{
\frac{\alpha^2 y}{2} -1 -s
}
$ \\ \hline
\end{tabular}
\end{center}
\vspace{5mm}
However the ground state solution of (\ref{eq:35}) in the Dirac case is 
\begin{eqnarray}
h(\rho) = F(0, 2 + 2s, \rho) = 1, \ \ \ \hbox{for} \ \ \ n=1.
\label{eq:36}
\end{eqnarray}
And both 
\begin{eqnarray}
\left ( \frac{\alpha E}{q} -1 -s \right ) = 0 
\ \ \ \hbox{and} \ \ \ \frac{1}{2} + s y =0
\label{eq:37}
\end{eqnarray}
holds and they give the same value of $q$, 
\begin{eqnarray*}
q= \alpha m.
\end{eqnarray*}
Therefore 
\begin{eqnarray*}
E &=&  \sqrt{m^2 (1-\alpha^2 )} \\
 & & \\ 
B &=& m-E = \left \{ 1 - \sqrt{1- \alpha^2} \right \} m.
\end{eqnarray*}

As is well-known, Dirac equation with Coulomb force can be solved 
for all cases of $n$ and $l$ in terms of two confluent hypergeomotric 
functions (multiplied by $\rho^s e^{-\rho/2}$).

However one cannot do 
\begin{eqnarray*}
N=0 \ \ \ \hbox{and} \ \ \ \delta =0
\end{eqnarray*}
simultaneously in the Breit case.
\begin{eqnarray*}
N=0 \ \ \ \hbox{means} \ \ \ \frac{\alpha^2 y}{2}-1-s =0.
\end{eqnarray*}
then 
\begin{eqnarray*}
\delta 
\ 
\lower.8ex\hbox{$\stackrel{\doteq}{ .}$}
\ 
\frac{1}{2} + s y 
\ 
\lower.8ex\hbox{$\stackrel{\doteq}{ .}$}
\ 
\frac{1}{4}.
\end{eqnarray*}
While $\delta = 0$ demands
\begin{eqnarray*}
\frac{1}{2} - \frac{\alpha^2}{8} y 
\ 
\lower.8ex\hbox{$\stackrel{\doteq}{ .}$}
\ 
0
\end{eqnarray*}
but 
\begin{eqnarray*}
\frac{\alpha^2 y}{2}
\ 
\lower.8ex\hbox{$\stackrel{\doteq}{ .}$}
\ 
2 \ \ \ \hbox{1st exiced state!}
\end{eqnarray*}
and $N$ cannot be zero.

From these situations, the Breit case is more complicated than 
the Dirac case, 
as described in this article.

\subsubsection{Confluent Heun function}

The solution $h(\rho)$ of differential equation (\ref{eq:14})
is confluent Heun function\cite{rf:5}, which has 5 parameters.
Unfortunately the author is not familier to this function, 
so that the auther has used physical ground and $1/y$ expansion of 
the wave function $h(\rho)$.

\subsection{$^{1}S_{0}$, $M \neq m$}
\label{sub:2-3}
Unequal mass case $M \neq m$ can be treated in exactly the same way as in 
\S\ref{sub:2-2} for the case of $M=m$.

The equation for the quantity $\tilde{h} (\rho)$:
\begin{eqnarray*}
F+K= \tilde{h}(\rho) e^{-\rho/2}
\end{eqnarray*}
is given by 
\begin{eqnarray}
\lefteqn{
\tilde{h}^{\prime \prime} + \tilde{h}^{\prime} \left [-1+ \frac{2}{\rho} 
- \frac{1}{\rho (1+y\rho)}
+ \frac{2(1+y\rho)}{\rho \{ (1+y\rho)^{2} - 
(\bar{M}-\bar{m})^{2} \rho^{2} \} } \right ]
} \nonumber \\
& & \qquad + \tilde{h} 
\left [- \frac{1}{\rho} + \frac{1}{2 \rho(1+y\rho)}-
\frac{1+y\rho}{\rho \{ (1+y\rho)^{2} - 
(\bar{M}-\bar{m})^{2} \rho^{2} \} } \right ]
\nonumber \\
& & \qquad + \tilde{h} \left [ \frac{1}{4} + 
\frac{\alpha^{2}}{4} \left \{ 1- 
\frac{(\bar{M}+\bar{m})^{2}\rho^{2}}{(1+y\rho)^{2}} \right\} 
\left \{ 
\frac{(1+y\rho)^{2}}{\rho^{2}}
-(\bar{M}-\bar{m})^{2}  \right \} \right ] = 0,
\label{eq:38}
\end{eqnarray}
where $\bar{M} = M/2 \alpha q$ and $\bar{m} = m/2 \alpha q$.  

As in \S\ref{sub:2-2}, $\rho=0$ and $\rho = \infty$ in eq.~(\ref{eq:38}) 
are the regular and irregular singular points, respectively.  
Therefore, we can adopt the same procedure as in \S\ref{sub:2-2}.

To remove the $1/\rho^2$ terms, put 
\begin{eqnarray*}
\tilde{h}(\rho) = \rho^s h(\rho), 
\end{eqnarray*}
finding again $s= -1 + \sqrt{1- \alpha^2 / 4}$.
The differential equation for $h(\rho)$ reads
\begin{eqnarray}
 & h'' &+ h' 
\left [ 
  -1 + \frac{\gamma}{\rho} 
  + \frac{1}{(1 + y \rho)^2 - (\bar{m} - \bar{M})^2 \rho^2 }
  \left \{ 
  \frac{(\bar{m}-\bar{M})^2 \rho}{1+ y \rho} + \frac{1 + y \rho}{\rho}
  \right \}
\right ] \nonumber \\ 
 & &+ h 
\left [ \left [
 \left ( \frac{\alpha^2 y}{2} -1-s  \right ) \frac{1}{\rho} 
\right.  \right.
\nonumber \\  
(A)\cdots& &+ \left [
   \left \{ 
    \frac{1}{(1+y\rho)^2 - (\bar{m}-\bar{M})^2 \rho^2} 
     \left ( 
      \frac{(\bar{m} - \bar{M})^2 \rho}{1 + y \rho} + 
      \frac{1 + y \rho}{\rho}
     \right )
   \right \} \frac{s}{\rho} - \frac{s}{\rho^2}
  \right ] 
\nonumber \\ 
(B)\cdots& &- \frac{1}{2} 
 \left \{ 
  \frac{1}{(1+y\rho)^2 - (\bar{m}-\bar{M})^2 \rho^2}
   \left ( 
    \frac{(\bar{m}-\bar{M})^2 \rho}{1 + y \rho} + 
    \frac{1 + y \rho}{\rho}
   \right )
  \right \}  \nonumber \\ 
(C)\cdots& &+ \frac{1}{4} + \frac{\alpha^2}{4} 
   \left \{ 
    y^2 - 2 (\bar{m}^2 + \bar{M}^2)
   \right \}
\nonumber \\ 
(D)\cdots& &+
\left. \left.
 \frac{\alpha^2}{4} 
\frac{(\bar{m}^2 - \bar{M}^2)^2}{(1+ y \rho)^2} \rho^2 
\right ] \right ] =0.
\label{eq:39}
\end{eqnarray}
We should notice here that (A)+(B) and (C) reduce
\begin{eqnarray*}
(A)+(B) &\rightarrow& \frac{-\delta}{\rho (1 + y \rho)}, 
\ \ \ \delta= \frac{1}{2}+ sy, \\ 
 & & \\ 
(C) &\rightarrow& 0, \ \ \hbox{and} \ \ (D) = 0, 
\end{eqnarray*}
when the equal masses case, $M=m$. 
(\ref{eq:39}) reduces to (\ref{eq:14}) for the case, 
$^{1}S_0, m=M$.

We shall rewrite the term (D) as follows:
\begin{eqnarray*}
         & (D) =& \frac{\alpha^2}{4} \frac{(\bar{m}^2 - \bar{M}^2)^2}{y^2} 
             \left ( \frac{y \rho}{1 + y \rho}  \right )^2 \\ 
(E)\cdots&     =& \frac{\alpha^2}{4} 
               \left ( 
               \frac{(\bar{m}^2 - \bar{M}^2)^2}{y^2} 
               \right )
             \left [ \left ( \frac{y \rho}{1 + y \rho}  \right )^2 
                     + \frac{2y}{\rho}
             \right ] \\ 
(F)\cdots&     & - \frac{\alpha^2 y}{2} 
              \left ( \frac{(\bar{m}^2 - \bar{M}^2)}{y^2} \right )^2
              \frac{1}{\rho}.
\end{eqnarray*}
The last term $(F)$ shall be added to the first line of th coefficient 
$h(\rho)$ in (\ref{eq:19}):
\begin{eqnarray}
h'' &+& h' \left\{ -1 + \frac{\gamma}{\rho} -2 (B) \right \} \nonumber \\ 
&+& h \left[ 
           \frac{\alpha^2 y}{2} 
            \left \{
                       1- 
                \left( 
                 \frac{\bar{m}^2 - \bar{M}^2}{y^2}
                \right )^2
            \right\} -1-s
    \right ] \frac{1}{\rho} \nonumber \\ 
&+& h \left [ (A)+(B)+(C)+(E) \right ] = 0.
\label{eq:40}
\end{eqnarray}
(A), (B), (C) and (E) are of the order of $1/y$ or $\alpha^2$. 
Now we can use the same technique described in \S\ref{sub:2-3}, 
$^{1}S$, $m=M$. 
\begin{eqnarray}
h(\rho) &=& F(-N, \gamma, \rho) + f(\rho),  
\label{eq:41}
\end{eqnarray}
\begin{eqnarray}
N &=& \frac{\alpha^2 y}{2} 
      \left \{ 
       1- \left ( \frac{\bar{m}^2 - \bar{M}^2}{y^2} \right )^2
      \right \} -1-s, \nonumber \\ 
N &=& n-1, \ \ \ n=1, 2, \cdots.
\label{eq:42}
\end{eqnarray}
$n$ is the principal quantun number. 
The background part $f$ is $O(\alpha^2)$ and can be found by solving 
\begin{eqnarray*}
f'' + f' \left \{ -1 + \frac{\gamma}{\rho} \right \}
+ f \left ( \frac{N}{\rho} \right ) 
= \left \{ 
+ 2 (B) \frac{d}{d \rho} -(A)-(B)-(C)- (E)
\right \} F(-N, \gamma, \rho).
\end{eqnarray*}
Thus the accuracy of the solution $F+f$ 
is quite good, terms of the order of $\alpha^4$ are ignored.

Other radial wave functions, $F, K, G, H, \tilde{G}$ and $\tilde{H}$ 
can now be writen down. 

The energy eigenvalues from (\ref{eq:42}) 
\begin{eqnarray}
\frac{\alpha^2 y}{2} 
 \left \{
  1- \left ( \frac{\bar{m}^2 - \bar{M}^2}{y^2} \right )^2
 \right \} 
&=& 
\frac{\alpha E}{4q} 
 \left \{
  1- \left ( \frac{m^2 - M^2}{E^2} \right )^2
 \right \} = n + s \equiv \bar{n}, \nonumber \\ 
 & & \nonumber \\ 
 E &=& \sqrt{M^2 - q^2} + \sqrt{m^2 - q^2}
\label{eq:43}
\end{eqnarray}
is easily calculated. 
Therefor the binding energy $B=M+m-E$ is 
\begin{eqnarray*}
B = M + m - E 
= \sum_{m=1}^{\infty} e_{m} \left (\frac{\alpha}{2 \bar{n}} \right )^{2m},
\end{eqnarray*}
\begin{eqnarray}
\left.
\begin{array}{l}
\displaystyle{
e_{1} = 
\frac{2mM}{(M+m)},
} \\
 \\ 
\displaystyle{
e_{2} = \frac{2mM}{(M+m)} \left \{ 1- \frac{3mM}{(M+m)^{2}} \right \},
} \\
 \\ 
\displaystyle{
e_{3} = \frac{4mM}{(M+m)} \left \{ 1- \frac{5mM}{(M+m)^{2}} + 
 \frac{5(mM)^{2}}{(M+m)^{4}} \right \}
}
\end{array}
\right \}
\label{eq:44}
\end{eqnarray}
and so on.
The first term is exactly the well-known form of the reduced mass.

It is easy to see that (\ref{eq:43}) reduces to (\ref{eq:23}) when $M=m$.

\vspace{10mm}

Knowing $F + K =e^{-\rho/2} \rho^s h(\rho)$, all relevant radial wave 
functions, $F, K, G, H, \tilde{G}$ and $\tilde{H}$ can easily be derived.


\begin{thebibliography}{99}

\bibitem{rf:1} G. Breit Phys.~Rev.~\underline{34} (1929) 553, 
\underline{36} (1930) 383, \underline{39} (1932) 616.

\bibitem{rf:2} H.A. Bethe and E.E. Salpeter, 
Quantum Mechanics of One- and Two-Electron Systems, 
p. 88-436, Bd XXXV Atom I, Handbuch der Physik, ed.~by S. Fl\"{u}gge 
(1957), esp.~p. 175-203 and p. 256-290.

\bibitem{rf:3} J.R. Sapirstein and D. R. Yennie, 
Theory of the Hydrogentic Bound States, p. 560-672, 
and V.W. Hughes and G. Zu Putlitz. Muonium, p. 827-904, 
esp.~ II.~2 Breit equation, p. 826.  Both in Quantum Electrodynamics, 
ed.~by T. Kinoshita, (Advanced Series on Direcctions in High Energy Physics, 
vol.~7), 1990 World Scientific.

\bibitem{rf:4} C. G. Darwin, Proc. Roy. Soc. London(A) \underline{115} 
(1928) 654. 
W. Gordon, Zs. f. Phys. \underline{48} (1928) 11.

\bibitem{rf:5} Prof. T. Rijken was kindly informed me of this fact 
(according to his son).



\end{thebibliography}
\end{document}